# CRYPTOGRAPHY FROM QUANTUM MECHANICAL VIEWPOINT

Minal Lopes[1] and Dr. Nisha Sarwade[2]

[1]Department of Electrical Engineering, VJTI, Mumbai, India
[2]Department of Electrical Engineering, VJTI, Mumbai, India

## ABSTRACT

*Cryptography is an art and science of secure communication. Here the sender and receiver are guaranteed the security through encryption of their data, with the help of a common key. Both the parties should agree on this key prior to communication. The cryptographic systems which perform these tasks are designed to keep the key secret while assuming that the algorithm used for encryption and decryption is public. Thus key exchange is a very sensitive issue. In modern cryptographic algorithms this security is based on the mathematical complexity of the algorithm. But quantum computation is expected to revolutionize computing paradigm in near future. This presents a challenge amongst the researchers to develop new cryptographic techniques that can survive the quantum computing era. This paper reviews the radical use of quantum mechanics for cryptography.*

## 1. INTRODUCTION

In 1984, Bennett (from IBM Research) and Brassard (from University of de Montreal) presented an idea of using quantum mechanics for cryptography (Quantum Cryptography). They proposed an algorithm (BB84) based on Heisenberg's 'uncertainty principle' for exchanging the *key* (QKD) along with a classical one-time pad (OTP) for information exchange [1]. After BB84, almost a decade later in 1991, Ekert reconsidered QC but with another unique principle of quantum mechanics known as 'Quantum Entanglement' [2]. Since these two base papers, many researchers have contributed towards QC protocols. All these protocols can be briefly divided in two categories, protocols based on 'Uncertainty principle' and based on 'quantum Entanglement' [3].

Though there is ample contribution towards development of protocols, the practical implementation of these protocols rather faces much more complications which make QC vulnerable. The Photon number splitting (PNS) attacks [4], a side channel attack [5], and imperfections in detectors [6] shows that QKD device imperfections need to be addressed. This gives rise to new category of protocols known as Device Independent (DI) QKD and Measurement Device Independent (MDI) QKD protocols.

The focus of this paper is on basic understanding of Quantum Cryptography and the underlying quantum principles while traversing along the roadmap of major QC protocol design evolution. The organization of this paper is as follows: In section II, we will discuss fundamentals of QKD and basic protocols based on Heisenberg's uncertainty principle. Section III will brief on evolutionary protocols based on Bell's theorem and quantum entanglement. In Section IV we will discuss the need for advanced protocol design and its approach using the principle of non locality and no signalling principle. We will also discuss the modern secure QKD protocol design through an example.

## 2. BASIC QKD PROTOCOLS

### 2.1. Fundamentals of QKD

The basic model of QKD protocol is shown in fig.1. It includes two classical users Alice and Bob, connected through quantum and classical communication channel, wishing to communicate securely. The quantum channel is used to exchange the *key*, whereas classical channel is used for actual information transfer.

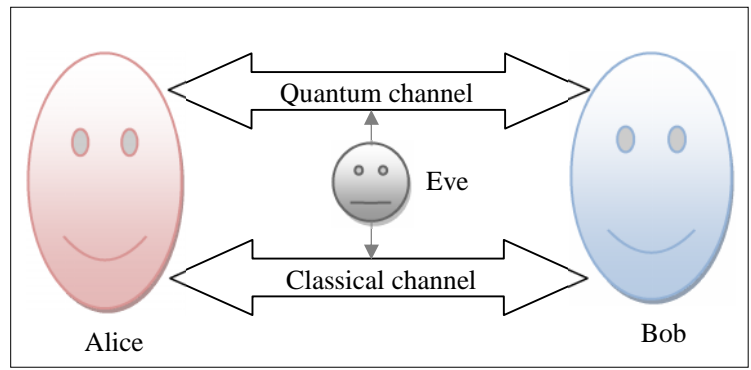

Figure 1.  Basic QKD Model

Figure shows Eve as a malicious eavesdropper, who have access to both the channels and is assumed to have infinite resources. Having this model established let us discuss the general steps involved in any QKD protocol. A typical QKD protocol comprises eight stages [7]:

1.  Random number generation by Alice(Sender)
2.  Quantum communication
3.  Sifting (basis reconciliation)
4.  Reconciliation
5.  Estimation of Eve's partial information gain
6.  Privacy amplification
7.  Authentication of public messages
8.  Key confirmation.

First, Alice (the sender) generates a sequence of random numbers from a hardware or software random number generator. Then, using the specified QKD protocol she encodes these random bits into the quantum states of a sequence of signals from her quantum light source and sends them over a "quantum channel" to Bob (the receiver). Bob applies a quantum measurement to each received signal and assigns it a bit value.

Next, Bob informs Alice over a classical channel, in which time slots he detected photons, but without revealing the bit value he assigned to each one. The bit strings corresponding to the signals detected by Bob are known as *raw keys*. Then, Alice and Bob select by public discussion a random portion of their raw keys, known as *sifted keys*. In an ideal system Alice and Bob's sifted key bits would be perfectly correlated.

In practice, Bob's sifted key is not perfectly correlated with Alice's. It contains transmission errors arising from background photons, detector noise and polarization imperfections. A typical error rate of 1 to 5 percent is presented by various literatures [8]-[10]. These errors must be located and corrected. Bob reconciles his sifted key with Alice's using an error correction method over their public channel, during which parity information about the sifted key is leaked. Their perfectly correlated reconciled keys are only partially secret now.

From the number of errors that Alice and Bob find in Bob's sifted key they are able to estimate an upper bound on any partial information that Eve might have been able to obtain on Alice's transmitted bit string. Quantum mechanics ensures that Eve's measurements would introduce a disturbance (errors) into Bob's sifted key that would be strongly correlated with Eve's partial information gain from them.

Alice and Bob extract from their reconciled keys a shorter, final bit string on which Eve's expected information is much less than one bit after an information-theoretic procedure known as "privacy amplification". In this procedure they use further public communications to agree to hash their reconciled keys into shorter final secret keys. For example, if Alice and Bob have 6 reconciled bits and their bound on Eve's information tells them that at most she knows 3 of these bits, they can agree to form two secret bits by XORing together the first 4 bits and the final 4 bits. Eve would have to guess at least one of the bits being XORed in each case and so would be ignorant of the outcome. These two bits are therefore suitable for use in a cryptographic key. More generally, Alice and Bob can form their final secret bits from the parities of random subsets of their reconciled bits [11].

## 2.2. The BB84 Protocol

BB84 is based on 'uncertainty principle' which states that unlike digital data the quantum data cannot be copied or measured without disturbing it. This immediately intimates that the eavesdropper cannot gain even a partial data without altering it. Also these alterations are most likely to be detected, solving the root cause problem of cryptography, known as 'eavesdropping'. This gives a radically different foundation for cryptography. In BB84, Alice generates a sequence of random numbers and encodes them using basis shown in Fig. 2. She chooses this basis randomly and transmits the polarized photon to Bob.

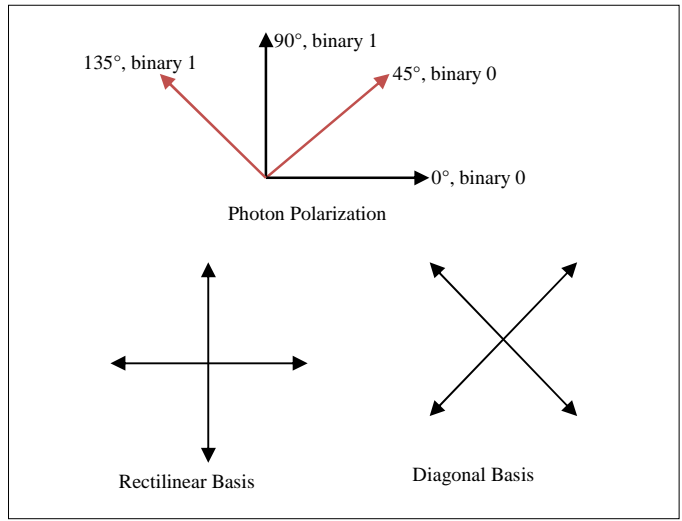

Figure 2. BB84 Encoding

Bob now measures the polarization of each received bit by choosing a random basis per bit. Now according to theory of probability, Alice and Bob are expected to choose same basis for more than 50% of cases on an average. Thus in an ideal case, with no eavesdropping and no transmission errors, the quantum bit error rate (QBER) of the transmission will be less than 50%. But Eve's presence will increase the QBER to more than 75% (ideally). This is illustrated in TABLE I and II through an example.

Table 1. Working of BB84 protocol without eavesdropping

| Alice's random bits | 1 | 0 | 1 | 1 | 1 | 0 | 1 | 1 | 1 | 0 | 1 | 0 | 1 | 1 | 0 | 0 | 0 | 1 | 1 | 1 | 1 | 0 | 1 | 0 | 1 | 1 | 0 | 1 | 0 |
| Alice's random Basis | + | × | + | + | × | + | × | × | × | + | + | × | + | × | + | × | × | + | + | + | × | × | + | × | × |  |  |  |  |
| Alice's photons with polarizations | ↕ | ⤢ | ↔ | ↕ | ⤡ | ↕ | ⤢ | ⤡ | ⤡ | ↔ | ↕ | ⤢ | ↕ | ⤢ | ↔ | ⤡ | ⤢ | ↕ | ↕ | ↔ | ⤡ | ⤢ | ↔ | ↕ | ⤢ |  |  |  |  |
| Assuming transmission errors at some random bits (e.g. bit number 4,24), error rate of 6.89% | | | | | | | | | | | | | | | | | | | | | | | | | | | | | |
| Bob's Random Basis | + | × | × | + | + | + | × | × | × | + | × | + | × | + | × | + | + | × | × | + | + | × | + | × | × | × | + |  |  |
| Bob's measurements (row key) | 1 | 0 | - | - | - | - | 1 | - | 1 | 0 | - | 0 | - | 1 | 0 | 0 | 0 | - | 1 | - | 1 | 0 | 1 | 1 | 1 | - | - | - | - |
| Bob reveals his basis for received bits to Alice on a classical channel | | | | | | | | | | | | | | | | | | | | | | | | | | | | | |
| Alice verifies the basis | C | C |  |  |  |  | C |  | C | C | W | C |  | C | C | C | C | W | C |  | C | C | C | C | C | C |  |  |  |
| Alice's sifted key | 1 |  |  |  |  |  | 1 |  | 1 | 0 |  | 0 |  | 1 | 0 | 0 | 0 |  | 1 |  | 1 | 0 | 1 | 0 | 1 |  |  |  |  |
| Bob's row key is correlated with Alice's for some bits and now include errors due to transmission impairments | | | | | | | | | | | | | | | | | | | | | | | | | | | | | |
| Bob's sifted key after public discussion | 1 | 0 |  |  |  |  | 1 |  | 1 | 0 |  | 0 |  | 1 | 0 | 0 | 0 |  | 1 |  | 1 | 0 | 1 | 1 | 1 |  |  |  |  |
| A simple parity check (reconciliation) will be used for rectifying this decorrelation | | | | | | | | | | | | | | | | | | | | | | | | | | | | | |

W - Wrong, C - Correct

From table I, it is clear that the QBER is less than 50% (13/29*100=44.82%, considering transmission errors), which is acceptable. Thus Alice and Bob can decide to continue the communication. In such case the sifted keys of both the parties will be partially correlated due to transmission errors. These errors can easily be removed by reconciliation process of 'parity check'. Table II depicts the case of presence of eavesdropper and transmission errors. Here the QBER is very high (24/29 *100 = 82.75% which is > 50%). This indicates the presence of eavesdropping in the communication. Both the parties can abort the transmission now.

BB84 have used quantum properties is such an efficient manner, that it has become the base for quantum cryptography. The disadvantage of BB84 is its assumptions for ideal quantum sources and detectors, which makes it vulnerable to PNS and detector noise attacks. Many other variants of it exist including B92 [12] using only two orthogonal states, SSP99 [13] using six states on three orthogonal bases. This means that an eavesdropper would have to choose the right basis from among 3 possibilities. This extra choice causes the eavesdropper to produce a higher rate of error thus becoming easier to detect.

Table 2. Working of BB84 protocol with eavesdropping and Transmission Errors

| Alice`s random bits | 1 | 0 | 1 | 1 | 1 | 0 | 1 | 1 | 1 | 0 | 1 | 0 | 1 | 1 | 0 | 0 | 0 | 1 | 1 | 1 | 1 | 0 | 1 | 0 | 1 | 1 | 0 | 1 | 0 |
|---|---|---|---|---|---|---|---|---|---|---|---|---|---|---|---|---|---|---|---|---|---|---|---|---|---|---|---|---|---|
| Alice`s random Basis | + | × | + | + | + | × | + | × | × | × | + | + | × | + | × | + | × | × | × | + | + | + | + | × | + | × | + | + | × |
| Alice`s photons with polarizations | ↕ | ⤢ | ↕ | ↕ | ↕ | ⤡ | ↔ | ⤡ | ⤡ | ⤡ | ↔ | ↕ | ⤢ | ↕ | ⤢ | ↕ | ↔ | ⤢ | ⤡ | ⤡ | ↕ | ↕ | ↔ | ⤡ | ↕ | ⤡ | ↔ | ↕ | ⤢ |
| Assuming transmission errors at some random bits (e.g. bit number 4,24) | | | | | | | | | | | | | | | | | | | | | | | | | | | | | |
| Eve`s random basis | × | + | × | + | + | + | × | + | × | + | × | × | + | + | + | × | + | × | + | × | × | + | + | + | × | × | × | + | × |
| Eve`s measurements | - | - | - | 0 | - | 0 | - | 1 | 1 | 0 | - | 0 | 1 | 1 | - | 0 | 0 | - | 1 | 1 | - | - | - | 1 | 1 | 1 | - | 1 | 0 |
| Eve`s bits (assuming random bits at nil measurements) | 1 | 1 | 0 | 0 | 1 | 0 | 0 | 1 | 1 | 0 | 1 | 0 | 1 | 1 | 1 | 0 | 0 | 1 | 1 | 1 | 0 | 1 | 0 | 1 | 1 | 1 | 1 | 1 | 0 |
| Eve`s random Basis | × | + | × | × | + | + | × | + | × | × | + | × | × | + | + | × | + | × | + | × | × | + | + | + | × | × | × | + | × |
| Eve`s Photons with polarizations | ⤡ | ↕ | ⤢ | ⤢ | ↕ | ↕ | ↔ | ↕ | ⤡ | ↔ | ↕ | ⤢ | ⤡ | ↕ | ↕ | ↔ | ⤢ | ⤢ | ↕ | ⤡ | ⤢ | ↕ | ↔ | ↕ | ⤢ | ⤡ | ⤢ | ↕ | ⤢ |
| Assuming transmission errors at some random bits (e.g. bit number 1,18) | | | | | | | | | | | | | | | | | | | | | | | | | | | | | |
| Bob`s Random Basis | + | + | + | × | × | × | + | + | × | × | + | × | + | × | + | + | × | × | × | + | + | × | × | + | × | + | × | × | + |
| Bob`s measurements (row key) | 1 | 1 | - | 0 | - | 0 | 0 | - | 1 | - | - | 0 | 1 | - | 1 | 0 | - | 0 | 1 | 1 | - | - | - | - | - | - | - | - | - |
| Bob revels his basis for received bits to Alice on a classical channel | | | | | | | | | | | | | | | | | | | | | | | | | | | | | |
| Alice verifies the basis | C | W | | W | | W | W | | C | | | C | W | | W | C | | | C | C | W | | | | | | | | |
| Alice`s sifted key | 1 | | | | | | | 1 | | | 0 | | | | | 0 | | | 1 | 1 | | | | | | | | | |
| Bob`s row key is correlated with Alice`s for some bits and now include errors due to transmission impairments | | | | | | | | | | | | | | | | | | | | | | | | | | | | | |
| Bob`s sifted key after public discussion | 1 | | | | | | | 1 | | | 0 | | | | | 0 | | | 0 | 1 | | | | | | | | | |
| Alice and Bob conclude that this transmission is not secure due to very high QBER (82.75%) and abort the transmission | | | | | | | | | | | | | | | | | | | | | | | | | | | | | |

W - Wrong, C - Correct

While there are number of other BB84 variants, one of particular interest was proposed in 2004 by Scarani, Acin, Ribordy, and Gisin named SARG04 [14] due to its capability of detecting the photon number splitting attack.

## 2.3. The SARG04 Protocol

This protocol is a modification of BB84 in sifting process. The bases used are non orthogonal states as shown in Fig (3), which cannot be discriminated deterministically. Also instead of revealing the bases, Alice announces the random states that she has used in preparing her photons. These random non orthogonal states can be defined as,

$$A_{w,w'} = \left\{ \left| wx \right\rangle, \left| w'z \right\rangle \right\}, \text{ with } w, w' \in \left\{ +, - \right\} \qquad (I)$$

Where, $\left| \pm x \right\rangle$ code for 0 and $\left| \pm z \right\rangle$ code for 1

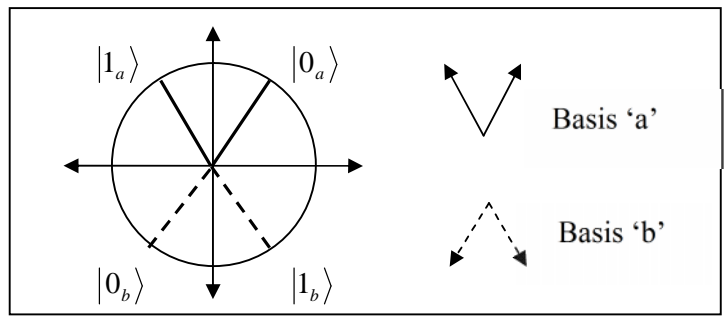

Figure 3. Two pairs of Non-orthogonal states in SARG04

The protocol is explained through an example in table III. Now when Alice sends these states, Bob will again measure the standard deviation against each polarization in that state. In any of the states if he gets the negative result, it indicates that Alice certainly has send the opposite state.

SARG04 has sifted key rate of 25% as compared to 50% for BB84. But it increases the attenuation for pulses with one and two photons, thus indicating the presence of eavesdropping in case of weak coherent pulses. This makes it robust against PNS attack. For definiteness, in Table 3 for bit 8, Alice announces the state as $A_{\cdots}$. When Bob measures his photon in both the polarizations, he gets negative result for one of the states ($\searrow$), indicating that Alice has send an opposite state ($\nwarrow$).

SARG04 has 'sifted key' rate of 25% as compared to 50% for BB84. But the attenuation increases specifically for pulses with one and two photons, thus indicating the presence of eavesdropping in case of weak coherent pulses. This makes it robust against PNS attack.

Table 3. Working of SARG04 protocol

| Alice's random bits | 1 | 0 | 1 | 1 | 1 | 0 | 1 | 1 | 1 | 0 | 1 | 0 | 1 | 1 | 0 | 0 | 0 | 1 | 1 | 1 |
|---|---|---|---|---|---|---|---|---|---|---|---|---|---|---|---|---|---|---|---|---|
| Alice's random Basis | | | | | | | | | | | | | | | | | | | | |
| Alice's photons with polarizations | | | | | | | | | | | | | | | | | | | | |
| Bob's Random Basis | | | | | | | | | | | | | | | | | | | | |
| Bob's measurements (row key) | | | | | | | | | | | | | | | | | | | | |
| Alice announces a random state | | | | | | | | | | | | | | | | | | | | |
| Bob's final measurements | | | | | | | | | | | | | | | | | | | | |

# 3. PROTOCOLS BASED ON QUANTUM ENTANGLEMENT (BELL'S THEOREM)

Entanglement is one of the most puzzling phenomena in quantum mechanics. It states that, if two particles are entangled, measuring the properties of one particle instantly determines the state of the second, no matter how far apart the two are [15]. Literature reveals possibilities of generating the pairs of such entangled photons [16-17], where each photon will always have a polarization orthogonal to the other photon. Also this polarization is indeterminate until a measurement is made. At the instant the measurement is made the polarization of the second photon becomes orthogonal to that of the first.

Ekert in his seminal paper [2] used this phenomenon for quantum cryptography by considering an un-trusted source of entangled photons between Alice and Bob. He claimed that there is no information carried by photons during transit, thus no eavesdropping. But if eavesdropper injects extra quanta (photon) in the communication, he suggested its detection by testing the Bell's Inequality at either ends.

## 3.1 Bell's Theorem and CHSH Inequality

Ekert's pioneering paper [2] used Bell's theorem for the first time to detect the eavesdropping. Although Bell's theorem was derived in view of EPR paradox for detection of local hidden variables, Ekert suggested its use to detect the presence of eve in the quantum communication.

Bell's theorem states that, "There exist predictions of quantum mechanics that are inconsistent with the predictions of local hidden variable theories [LHVT]". Bell [18] provided a mathematical description for local hidden variable theories, which are hypothetical physical theories, based on the EPR assumptions of local causality and physical reality. Bell's formalism allows us to test these conditions experimentally by violating some inequalities (Bell inequalities), which are satisfied by LHVT. Although there are various experiments proposed for testing Bell's Inequality, the most popular one is the CHSH inequality [19].

The simplified form of a CHSH inequality is,

$$S = E(a,b) + E(a,b^{'}) + E(a^{'},b) - E(a^{'},b^{'})$$

$$and \hspace{4cm} (II)$$

$$S \leq 2$$

Where, $a$ and $a^{'}$ are the detector settings on side A, $b$ and $b^{'}$ are the detector settings on side B. The four combinations of these settings are tested in separate sub-experiments. The terms $E(a,b)$ etc. are the quantum correlations defined as the expectation value of the product of the outcomes of the experiment. The mathematical formalism of quantum mechanics predicts a maximum value for 'S' as $2\sqrt{2}$ , which is greater than 2, and CHSH violations are therefore predicted if the correlations are purely quantum. This property is utilized to check if Eve malfunctions with the quantum transmission.

In 1992, Bennett, Brassard and Mermin [20] proved that E91 is the entanglement version of BB84 and any variant of BB84 could be adapted to use an entangled photon source instead of Alice being the source. Enzer et al [21] have described an entangled version of the SSP protocol with added security. The entanglement version of SARG04 was proposed by Fung et. al.[22]. Let us discuss the E91 protocol as a classical example of Entanglement based protocols.

### 3.2 E91 Protocol

Assuming a source (EPR) that emits pairs of spin-1/2 particles, in a singlet state, first Alice and Bob distribute $n$ pairs of qubits in a Bell states represented as,

$$\left|\psi^{+}\right\rangle = \frac{1}{\sqrt{2}}\left(\left|0\right\rangle_{A}\left|1\right\rangle_{B} + \left|1\right\rangle_{A}\left|0\right\rangle_{B}\right) \hspace{2cm} (III)$$

Equation (III) represents entangled state particles (The bell states can be represented in four different combinations of two qubits [23]) that fly apart towards Alice and Bob and can be prepared to have spin components along one of the three directions (x, y, z) given by unit vectors $a_i$ and $b_j$ $(i, j = 1,2,3)$ having angles, $\phi_1^a = 0, \phi_2^a = \frac{1}{4}\pi, \phi_3^a = \frac{1}{2}\pi$ and $\phi_1^b = \frac{1}{4}\pi, \phi_2^b = \frac{1}{2}\pi, \phi_3^b = \frac{3}{4}\pi$ , as shown in fig. 4. The subscripts "a" and "b" refer to Alice and Bob respectively.

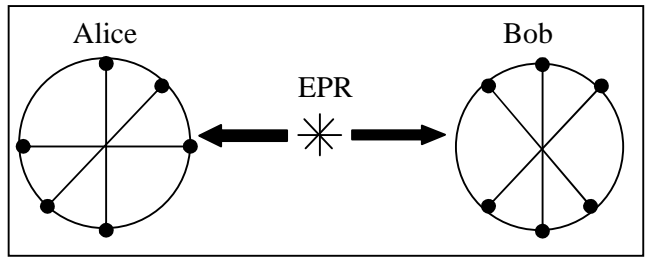

Figure 4. E91's basis used by Alice and Bob in normal phase [24]

For each instance $i = 1,2,3,......n$ , each user randomly and independently chooses the basis (analyzer's orientation) to perform the measurements. Later only those cases where Alice and Bob's measurements match are used, while others are discarded. Among the matched cases, Alice and Bob agree on some results to be used for checking phase while remaining cases being used for normal phase.

In normal phase, each user performs $S(z)$ measurements. The outcomes of these measurements are perfectly correlated and thus used as a key later. In the checking phase, each user performs the measurements for Bell's inequality violation of the state by using only two basis as shown in fig. 5. Alice (Bob) randomly and independently chooses one between the two directions $a$ and $a'$ ($b$ and $b'$) for measurements.

Then Alice (Bob) performs spin-measurements in the chosen direction. Spin-measurement in direction $p$ $(q)$ is denoted as $S(p)$ $(S(q))$ where $p = a, a'$ $(q = b, b')$. The probability that Alice gets a result $\pm 1$ in spin-measurement $S(p)$ and Bob gets a result $\pm 1$ in spin-measurement $S(q)$ is denoted as $P_{\pm\pm}(p, q)$. The correlation function $E(p, q)$ is given by

$$E(p,q) = P_{++}(p,q) + P_{--}(p,q) - P_{+-}(p,q) - P_{-+}(p,q) \qquad (IV)$$

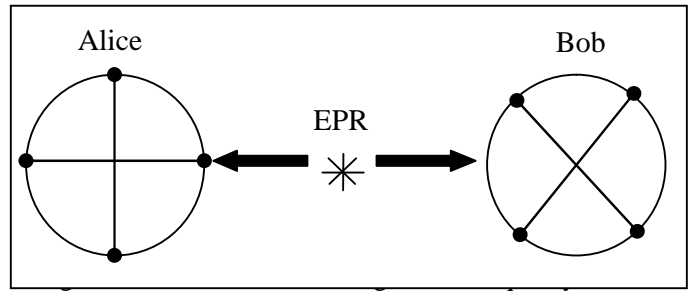

and the Bell's inequality is given by [25] (refer equation II)

$$\left| E(a,b) + E(a,b') + E(a',b) - E(a',b') \right| \le 2 \qquad (V)$$

If the particles are perfectly entangled then this inequality will produce the value of $2\sqrt{2}$, which is greater than 2, violating the inequality. If Eve provides separable states (non entangled) to Alice and Bob she will get detected in checking phase. If she provides perfectly entangled states, she may pass the checking phase but will not have any information about the key. Thus E91 shows the practical application of Bell's theorem in cryptography, which can test the safety of the key distribution.

## 4. NEXT GENERATION QKD

All though it is claimed [26-27] that the security of QC depends on the fundamental laws of quantum physics, the working of QC protocols is based on certain assumptions. Perfect device design (quantum sources and detectors), secure environment (physical location of legitimate parties) are some of them. Such assumptions leave QC vulnerable. As discussed earlier, the PNS attack exploits the replacement of single photon sources by weak coherent pulses source. The E91 assumes the detector efficiency, which if not 100% can be exploited to produce detection loopholes. The dark count rate of detectors further leads to reduced detection efficiency which is analogous to the losses in the quantum channel.

The solution over all these limitations is the motivation for next generation QKD protocols known as Device- Independent QKD (DIQKD) protocols. The idea of device independency was originally derived through E91 protocol, where the source is considered untrusted and violation of Bell's inequality serves to decide the security of protocol. But it is proved that the violation of Bell's inequality can be faked by using classical light [28] and utilizing the detection loophole in single photon detector [29]. Thus E91 is one step away from the actual DIQKD purpose. To achieve total device independence Alice and Bob should not only distrust the source, but should also assume that their measuring apparatuses are untrusted and controlled by Eve. We thus can define following set of assumptions as the ultimate aim of fully DIQKD protocols [30].

1.  *Alice and Bob's physical locations are secure.*
2.  *They have trusted random number generators.*
3.  *They have trusted classical devices to store and process the classical data.*
4.  *They share an open authenticated classical channel.*
5.  *Quantum physics is correct.*

The pioneering papers for DIQKD were proposed by Barrett et.al. (BHK05) [31] and Acin et.al. (AGM06)[32]. Till BHK05, the existing security proofs of QC [33-34] were based on the no-cloning theorem.

## 4.1 The No-cloning Theorem

The no-cloning theorem states that "it is impossible to produce a perfect copy of an unknown quantum state with any nonzero success probability".

BHK05 for the first time used quantum non locality and no-signalling principle for proving the security of QC.

## 4.2 Quantum Non-locality

It states that there exist some correlations that are non local and also follows the relativistic principle of causality. The popular example of such non local correlation is quantum Entanglement.

## 4.3 No-signalling Principle

The no-signalling principle states that communication between two distant parties cannot be performed without the transmission of any physical systems, despite any physical resources that the parties may share.

It can be shown that the quantum entanglement is consistent with relativity by satisfaction of the no-signalling principle [35]. Thus if an eavesdropper is limited by no-signalling principle, means he cannot prepare two or more physical systems in a joint state such that a local measurement on one system may transfer information to another distinct system, it is possible to develop fully device independent QKD protocols.

It was noted later that the physical implementation of BHK05 was not possible because of its stringent requirements about Alice and Bob's noise free channel. Also this protocol generates a single shared secret bit by making many uses of the channel reducing the key rate to inefficient value. The AGM06 was considered more practical DIQKD protocol.

### 4.4 The AGM06 Protocol

AGM06 is the modified E91 protocol as it uses Bell's inequality for proving the security. The strength of AGM06 is that it imparts no assumption on quantum devices (aiming towards fully device independence) and its security depends only on prerequisite fundamental assumptions discussed above.

As compared to E91, AGM06 expects Alice and Bob to share a quantum channel consisting of a source that emits pairs of photons in an entangled state. Alice chooses one out of three possible measurements $A_x (x = 0,1,2)$ to perform on her photon, while Bob chooses one out of two measurements $B_y (y = 1,2)$. The output of all measurements is the binary bits $a, b \in \{+1, -1\}$. The raw key is then extracted from outcomes of the measurement pair $\{A_0, B_1\}$. The quantum bit error rate is defined as $Q = P(a \neq b \mid x = 0, y = 1)$. '$Q$' estimates the number of correlations between Alice and Bob's symbols, and thus quantifies the amount of classical communications needed for error correction. The measurements $A_1, A_2, B_1$ and $B_2$ are used to estimate security by using the CHSH polynomial (refer equation (II)) as given below,

$$S \leq P(a = b \mid x = 1, y = 1) + P(a = b \mid x = 1, y = 2) + P(a = b \mid x = 2, y = 1) - P(a \neq b \mid x = 2, y = 2) \qquad (VI)$$

The violation of Equation (VI) implies that in three measurement choices, Alice and Bob are correlated, while in the fourth case they are anticorrelated. Here $S > 2$ is expected to ensure the security.

## 5. CONCLUSION

Quantum Cryptography provides a promising future in the field of data and network security through the fundamental quantum mechanical principles. This paper has presented a review of foundational QC protocols along with the modern quantum principles that aids cryptography. It is found that most protocols are based on BB84 and E91 and are vulnerable when implemented with realistic physical devices. It is also intimated that possibility of fully equipped and post quantum eavesdropper, puts strong limitations on feasibility of QC. Thus QC needs additional efforts and contribution from researchers in development of protocols and testing their security.

With the development in quantum communication, this field finds tremendous scope in near future. Therefore the implementation feasibility of QC without any assumptions needs to be tested. The new generation protocols like DIQKD are proved secure theoretically but lack the physical implementation. QC has still a long way to go in terms of mathematical modeling. The performance metrics like key rate, quantum bit error rate needs to be exploited first through mathematical and then physical modelling.

Quantum cryptography thus seems very fascinating field of research which needs collaborative efforts from diverse fields like physics, optics, computer science and mathematics. It alongside demands the development of various other fields like single photon sources, free space optical communication, optical detectors which in future will also be able to provide compatibility to the post quantum cryptographic research.

## Authors

Minal Lopes received B.E. and M.E. degrees in Electronics Engineering from Mumbai University. She is working as an Assistant Professor at St. Francis Institute of Technology Mumbai, Ind ia. Currently, she is pursuing PhD under guidance of Dr. Nisha Sarwade at the VJTI, Mumbai, India. Her areas of specialization are Quantum cryptography, Data and Network security and Next generation Networks.

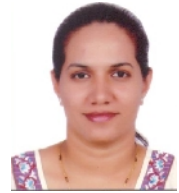

Dr. Nisha Sarwade received B.E. degree in Electronics Engineering from Jiwaji University, Gwalior and M.E. (Solid State Electronics) and PhD (Electronic Engineering) from University of Roorkee. She was working as a lecturer at the Universit y of Roorkee during 1983-1987. Currently she is working as an Associate Professor at the VJTI, Mumbai, India. Her research interests include Nano Electronics with emphasis on CNT, Compound semiconductors, High-k dielectrics and flash memories and Microwave circuit design. She has 56 national as well as international publications to her credit.

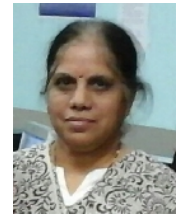